\documentclass[ras]{agu2001}
\usepackage{graphicx}

\authorrunninghead{BOWER}
\titlerunninghead{Extended Source RFI Mitigation}
\authoraddr{Geoffrey C. Bower,
Radio Astronomy Laboratory, UC Berkeley,
Berkeley, CA 94720.
(gbower@astro.berkeley.edu)}
\begin{document}

%% ------------------------------------------------------------------------ %%
%
%  IMAGE DISPLAY IN DRAFT MODE
%
%% ------------------------------------------------------------------------ %%

% UNCOMMENT THE FOLLOWING CODE IF YOU NEED TO INCLUDE IMAGES IN DRAFT MODE
% TO MAKE A PDF FOR FIRST SUBMISSION.
%
\setkeys{Gin}{draft=false}

% PLEASE NOTE: WE DO NOT WANT ANY GRAPHICS INCLUDED WHEN THE FINAL VERSION
% OF YOUR TEXT FILE IS SUBMITTED
% PLEASE COMMENT OUT ALL \includegraphics AND \figbox COMMANDS
% WHEN USING THE DRAFT MODE TO SUBMIT YOUR ACCEPTED ARTICLE
%
% For accepted papers
% (i) the graphics should not be included,
% (ii) figures and tables should be listed at the end of the file.
%

%% ------------------------------------------------------------------------ %%
%
%  TITLE
%
%% ------------------------------------------------------------------------ %%

\title{Radio Frequency Interference Mitigation for Detection of Extended Sources 
with an Interferometer}
%
% e.g., \title{Terrestrial Ring Current:
% Origin, Formation and Decay $\alpha\beta\Gamma\Delta$}

%% ------------------------------------------------------------------------ %%
%
%  AUTHORS AND AFFILIATIONS
%
%% ------------------------------------------------------------------------ %%

% Method 1 (for all journals, except Reviews of Geophysics, which
% should use method 3):
% For three or fewer author/affiliation blocks, use \author{} and \affil{}

\author{Geoffrey C. Bower}
\affil{Radio Astronomy Laboratory, UC Berkeley,
Berkeley, CA 94720, USA}

\begin{abstract}
Radio frequency interference (RFI) is a significant problem for current and future
radio telescopes.  We describe here a method for post-correlation cancellation of 
RFI for the special case of an extended source observed with an interferometer that spatially
resolves the astronomical signal.  In this circumstance, the astronomical signal is
detected through the auto-correlations of each antenna but is not present in the
cross-correlations between antennas.  We assume that the RFI is detected in both auto- and cross-correlations,
which is true for many cases.
The large number of cross-correlations can provide a very high interference to noise
ratio reference signal which can be adaptively subtracted from the autocorrelation
signals.  The residual signal is free of interference to significant levels.  We 
discuss the application of this technique for detection of the spin-flip transition
of interstellar deuterium with the Allen Telescope Array.  The technique may also be
of use for epoch of reionization experiments and with multi-beam feeds on
single dish telescopes.

\end{abstract}

\begin{article}

\section{Introduction}

Radio frequency interference (RFI) mitigation is fast becoming a necessary aspect of
radio astronomy as terrestrial and space-based radio transmitters become more
widespread and more powerful, as radio telescopes become more sensitive and search for
ever fainter signals, and as radio astronomers seek to observe outside of the
protected radio astronomy bands.  A variety of RFI mitigation 
methods have been developed that rely on a range of signal processing techniques 
\citep[e.g.,][]{2000ApJS..131..355L}.
Many of these techniques are generic in their application in the sense that
they are not specific to the scientific goals and the types of interferers.  There
is much to be gained, however, from techniques which are specific to a given type of
astronomical observation or a given type of interferer 
\citep[e.g.,][]{2001ApJS..135...87E}.

We describe here a technique developed for detection of a diffuse astronomical source
with a high resolution interferometer in the presence of point-like or partially-resolved
interference.  The technique
exploits the fact that the astronomical signal is detectable only through the auto-correlation
of each antenna's signal, while the interference is detected in both the auto- and cross-correlations
of the signals.  For an interferometer with even a modest number of elements, 
the number of cross-correlations 
is significantly greater than the number of auto-correlations.  Thus, the problem becomes
similar to a reference antenna method with multiple reference antennas.

In \S 2, we describe the technique and show results of computer simulations which demonstrate
its performance characteristics.  In \S 3, we discuss possible applications of this technique.
These include detection of the interstellar deuterium spin-flip transition
with an interferometer with baselines on the order of
100 meters, and detection of the cosmological epoch of reionization signature.  The method is 
also applicable for multi-beam feeds on single-dish radio telescopes.

\section{Cross-correlation Subtraction}

We consider our method for an interferometer with $N_a$ elements configured in a non-regular
pattern.  Each element receives signal from the sky and converts the signal to baseband.
The baseband signals are processed by a correlator which produces auto-correlation and cross-correlation
power spectra with $N_{ch}$ channels.  These spectra are computed from $N_s$ digital samples.

We characterize the signal received at each antenna $i$ as the sum of the astronomical
signal, $S_i$, the receiver noise, $N_i$, and the interference, $I_i$:
\begin{equation}
E_i = S_i + N_i + I_i.
\end{equation}
The auto- and cross-correlation power spectra are then
\begin{eqnarray}
<E_i E_i^*> = <S_i S_i^* > + <I_i I_i^* > + < N_i N_i^*>, \\
<E_i E_j^*> = <S_i S_j^* > + <I_i I_j^* > + < N_i N_j^*>.
\end{eqnarray}

We assume that the source term in the cross-correlation function $<S_iS_j^*>$
goes to zero while the source term in the auto-correlation function $<S_iS_i^*>$
is non-zero.  This assumptions holds for the case of an extended, smoothly distributed
brightness distribution observed with a high resolution interferometer.  
In particular, this condition holds for baselines with a fringe spacing much smaller
than the angle subtended by the source on the sky.
The difference between the auto- and cross-correlation functions is then:
\begin{eqnarray}
<E_i E_i^*>  -  <E_i E_j^*>= <S_i S_i^* > + \nonumber \\
<I_i I_i^* > - <I_i I_j^*> + < N_i N_i^*> + < N_i N_j^*> 
\end{eqnarray}
Under the assumption that the interference is not resolved by the interferometer (i.e.,
that it is a point source), the second and third terms on the right hand side of the
equation sum to zero.  We are then left with a signal that only includes the effects of
the astronomical source and the system noise.  The system noise term will be dominated
by the noise auto-correlation and we are left with a residual signal:
\begin{equation}
<E_i E_i^*>  -  <E_i E_j^*>= <S_i S_i^* > + < N_i N_i^*>,
\end{equation}
which is simply the auto-correlation signal in the absence of interference.

The cancellation of interference that takes place between the auto- and cross-correlation
signals as described above requires an idealized system.  Under realistic conditions, a number of effects
will limit the difference between the auto- and cross-correlation of the interference.
These include multiple sources of interference, multi-path propagation of interference, and
variable antenna gain between two different array elements in the direction of the
interferer or interferers.  Adaptive cancellation methods introduce new degrees of
freedom that make the realistic problem tractable.
We suggest a method that minimizes the residual $\epsilon$ through the adjustment of
the cross-correlation gains $\mu_{ij}$, where
\begin{equation}
\epsilon = ( \sum_i <E_i E_i^*> - \sum_{i} \sum_{j>i} \mu_{ij} < E_i E_j^* > )^2.
\label{eqn:epsilon}
\end{equation}

This method is equivalent to a reference antenna adaptive cancellation
method in which the astronomy signal is the sum of the auto-correlations and the
reference signal is the weighted sum of the cross-correlations.
We can use the results for the reference antenna adaptive cancellation technique
to estimate the effectiveness of this methodology
\citep{1998AJ....116.2598B}.
The attenuation of interference
in the astronomy signal goes as
\begin{equation}
A = ( {\rm INR_{ref} } + 1)^2,
\end{equation}
where ${\rm INR_{ref}}$ is the interference power to noise power ratio (INR)
in the combined reference signal.
Thus, the residual astronomy signal will have interference at the level of 
\begin{equation}
{\rm INR_{resid}} = A^{-1} {\rm INR_{ast}},
\end{equation}
where ${\rm INR_{ast}}$ is the initial INR in the sum of
the auto-correlations.

We can determine the ${\rm INR}$s for the reference and astronomy signals as follows.  We assume
that the signal is Nyquist-sampled at a rate $f$ for $N_s$ samples.  
The ${\rm INR}$ in the autocorrelation of a single sample is ${\rm INR}_0$.  In the
case of the astronomy signal, the astronomy signal grows coherently with number of antennas
and number of samples while the noise grows incoherently leading to:
\begin{equation}
{\rm INR_{ast}} = \sqrt{ N_a N_s} {\rm INR_0}.
\end{equation}
In the reference signal, the interference and noise signals grow similarly; however,
the number of signals contributing to each sum grows as $N_a(N_a-1)$.  This leads to 
\begin{equation}
{\rm INR_{ref}}= \sqrt{ N_a (N_a -1) /2 N_s} {\rm INR}_0.
\label{eqn:inrref}
\end{equation}
Substituting these results into Equation 8, we find
\begin{equation}
{\rm INR_{resid}} = { \sqrt{ N_a N_s} {\rm INR_0} \over 
( \sqrt{ N_a (N_a -1) /2 N_s }{\rm INR}_0 + 1)^2 }.
\label{eqn:inrresid}
\end{equation}
In the limit of $\sqrt{ N_a (N_a -1) /2 N_s} {\rm INR}_0 >>1$ and $N_a >> 1$, we find that 
${\rm INR_{resid}} \approx N_a^{-3/2} N_s^{-1/2} {\rm INR_0}^{-1}$.

We can also compute the ratio of signal to interference (${\rm SIR}$).  The signal to noise ratio 
in the autocorrelation sum is 
\begin{equation}
{\rm SNR_{ast}} = \sqrt{ N_a N_s} {\rm SNR_0},
\label{eqn:snrast}
\end{equation}
where ${\rm SNR_0}$ is the signal to noise ratio of a single auto-correlation sample.
Thus, ${\rm SIR}$ is the ratio of Equations \ref{eqn:snrast} to \ref{eqn:inrresid}.
In the high ${\rm INR_0}$ limit discussed 
above, ${\rm SIR} \approx N_a^2 N_s {\rm SNR_0} {\rm INR_0}$.

\begin{figure}
 \noindent\includegraphics[width=20pc]{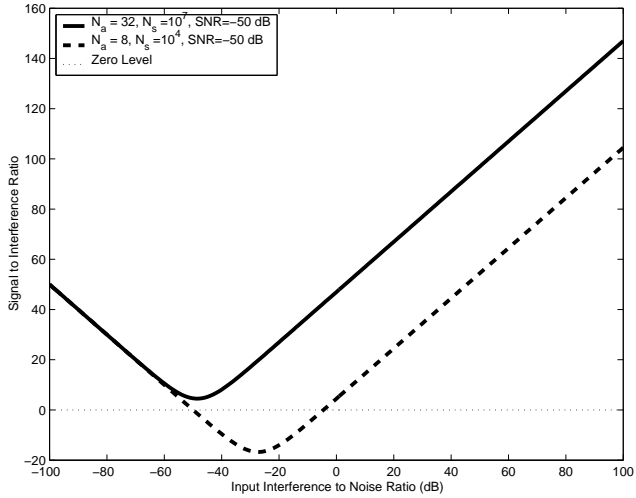}
 \caption{Signal to interference ratio in the residual astronomy signal as a function
of initial interference to noise ratio for two cases.  In both cases, the signal is
assumed to be 50 dB below the noise.  In case 1 (solid line), with 32 antennas and
$10^7$ samples per iteration the signal is always greater than the interference in
the residual.  In case 2 (dashed line), with 8 antennas and $10^4$ samples per iteration, 
weak interference can corrupt the signal.
\label{fig:sir}}
\end{figure}

We plot the SIR in Figure~\ref{fig:sir} for two cases.  We assume ${\rm SNR}_0=-50$ dB.
In both cases the SIR increases
dramatically for very large ${\rm INR}_0$.  That is, powerful interferers are relatively easy to
detect and subtract.  Both cases also show an inflection point at which the SIR reaches
a minimum.  Below this point, the interference is too weak to be detected and cannot be
cancelled.  If the minimum occurs at ${\rm SIR} > 0$, then the technique has succeeded
in protecting the signal against all interferers.  In our example, 
this holds for the case with large $N_a$ and $N_s$.
For a smaller $N_a$ and $N_s$, the ${\rm SIR}$ dips below zero, leaving the observations
potentially corrupted by weak interference.

\subsection{Simulations and Practical Considerations}

We describe a method for solving for the parameters $\mu_{ij}$ described
in Equation \ref{eqn:epsilon}.  We use a Wiener method to calculate the
best values for $\mu_{ij}$ but other methods are available
\citep{2000Vas}.  The Wiener
method has the advantage that it directly calculates the values; however, it 
is computationally expensive and may be replaced in practice by
an iterative method \citep{2001ApJS..135...87E}.

We recast the problem into a single astronomical signal $\hat{\bf s}$, which
is the sum of the auto-correlation functions above, and $N_b = N_a(N_a-1)/2$ 
reference
antennas $\hat{\bf R}$.  Both $\hat{\bf s}$ and $\hat{\bf R}$ are functions of
channel number.  We also add an additional reference signal which is a 
constant as a function of channel number to remove the effects of the
bandpass.  For more complex bandpasses, multiple template functions such
as a tilt or a ripple can be included.  It is important that $N_{ch} > 
N_{ref}$, where $N_{ref}$ is the number of reference signals.
If not, the solution is over-determined and all signal will be removed
by the algorithm.  The Wiener solution is then
\begin{equation} 
\hat{\bf \mu} = (\hat{\bf R} \hat{\bf R'} )^{-1} \hat{\bf R} \hat{\bf s}.
\end{equation}
 
\begin{figure}
 \noindent\includegraphics[width=20pc]{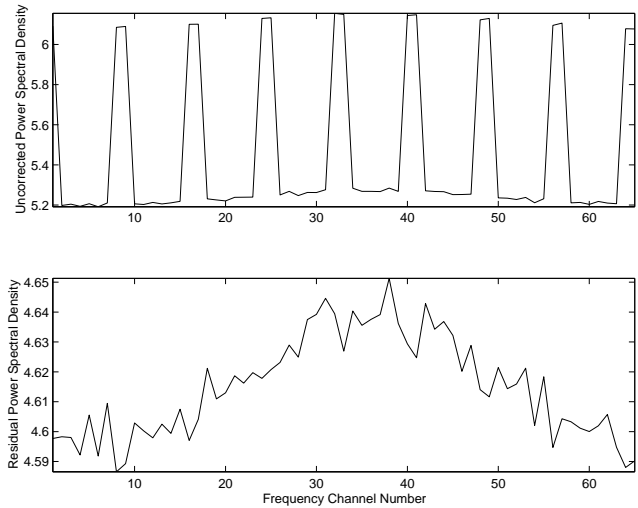}
 \caption{Results of a simulation of the cross-correlation subtraction 
technique.  The
top panel shows the astronomy signal without interference reduction.  The
bottom panel shows the astronomy signal with interference reduction.  Details
of the simulation are given in the text.
\label{fig:resid}}
\end{figure}

In Figure~\ref{fig:resid}, we show the result of a simulation of 
the cross-correlation method.  We used an 8-element array with 65
spectral channels.  The interference was a frequency comb with a spacing
of 8 channels and a width of 2 channels in each peak.  The gain
of each antenna was modulated in each iteration separately for the interferer
and for the source.  The interference power was set to 20\% of the
noise power.  The signal
was characterized as a Gaussian with width 16 channels with a peak power that
is 1\% of the noise power.  Each iteration consisted of $100$ samples.  We
iterated for 2000 iterations.  For 10 kHz of channel bandwidth, this would
correspond to a 0.3 second integration.  The interference is cleanly
removed and the weaker signal is readily apparent in the residual 
spectrum with an amplitude of 1\% of the noise power.  The expected attenuation
of the interference in the residual signal is 21 dB.
The actual reduction of
interference is $\sim 20$ dB, comparable to that expected.

There are a few drawbacks and additional considerations
to the method outlined here.  The first problem is
that the mean power off source
of the residual signal is reduced from the no-interference
case.  This may lead to calibration uncertainty.  This effect results from
the fact that all of the reference signals are positive quantities.  The
second problem is that in cases where the astronomical source signal is strong, the interference
is over-subtracted.  Since the algorithm is attempting to minimize the total
spectrum, the presence of an astronomical signal biases the subtraction.  
The algorithm will also remove strong astronomical point sources
from the residual signal.  Whether this is a problem is a function of the
particular experiment being undertaken.  Additionally, the effects of partial resolution
of the astronomical source of interest have not been fully explored.  
In what SNR regime does the remaining component of the source serve to mitigate
the source itself?
Finally,
interferers in the near-field as well as those undergoing significant multi-path
propagation may be resolved by the interferometer.  It may not be possible to
remove interferers of this kind from the astronomy signal.

Alternative methods of performing the subtraction may avoid some of these problems.  
Knowledge of the noise power spectrum obtained through careful calibration, 
for instance, would allow one to give
each spectrum zero mean.  One could also generalize the two-reference antenna
method described in \citet{2000AJ....120.3351B} for this multiple reference antenna case.
It is also worthwhile to explore the effect of using subsets of baselines to
generate the reference signal.

\section{Applications}

\subsection{Interstellar Deuterium}

The 92 cm deuterium spin flip transition (DI) is one of the most important radio spectroscopic
lines not yet detected 
%(Weinreb 1962, Anantharamiah \& Radhakrishnan 1979, Blitz \& Heiles 1987, Chengular 1997).  
\citep{1962ApJ...136.1149W,1979A&A....79L...9A,1987ApJ...313L..95B,1997A&A...318L..35C}.
The transition is the deuterium analog of
the hydrogen 21 cm line (HI) and arises from a flip in the spin direction of the electron with
respect to the nuclear spin.  The deuterium to hydrogen ratio ($N(D)/N(H)$)
which can be determined from detection of the DI line 
is an important constraint
on cosmic nucleosynthesis.  

The signal is known to be quite weak.
\citet{1987ApJ...313L..95B}
place an upper limit of $N(D)/N(H) < 3 \times 10^{-5}$,
which is consistent with ultra-violet detections at $2 \times 10^{-5}$.   This ratio
implies a line strength towards the Galactic anti-center on the order of 1 mK, on
the order of -50 dB times the system temperature of a radio telescope observing
in the Galactic plane at these wavelengths.  Since the DI emission traces the HI
emission, the line width is expected to be quite narrow ($\sim 10$ km/s)
towards the anti-center.

\begin{figure}
 \noindent\includegraphics[width=20pc]{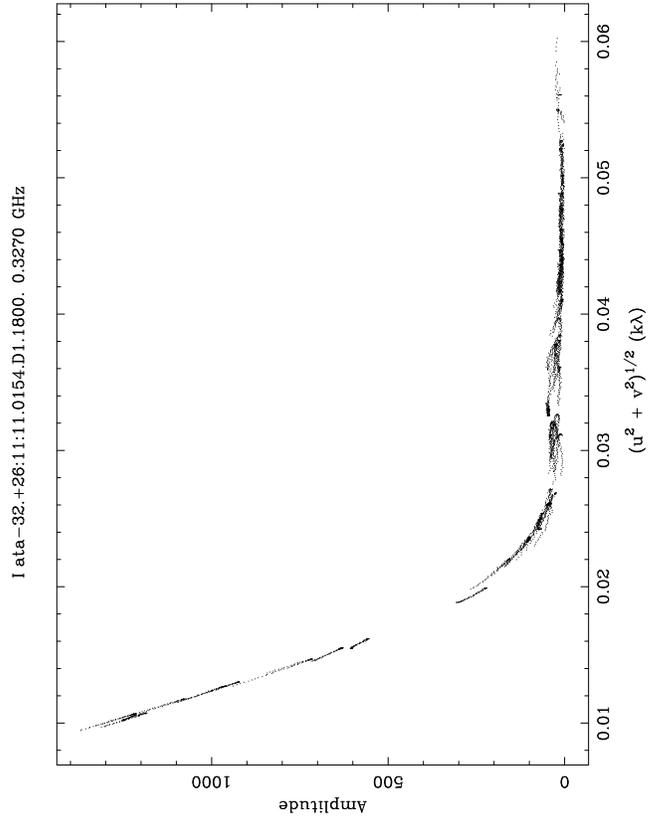}
 \caption{
Correlation amplitude as a function of baseline length for deuterium as observed with
the ATA-32.  The signal drops off sharply with increasing baseline length making this
problem well-suited to the cross-correlation subtraction method.
\label{fig:uvd}}
\end{figure}

We have simulated DI observations with the 32 element configuration of the Allen
Telescope Array \citep{2004SPIE.5489.1021D}.  The ATA-32 employs 6.1m paraboloids distributed in two dimensions
with baseline lengths that range from 8m to $\sim 100$m.  We simulated a snap shot
observation towards the Galactic anti-Center using MIRIAD software.  The sky model assumes
that the DI traces HI as observed by the Leiden Dwingeloo survey 
\citep{1997agnh.book.....H}.  We determined
the correlated amplitude as a function of baseline length (Figure~\ref{fig:uvd}).  The
source is substantially resolved on baselines longer than about 20m.  These
baselines comprise more than 90\% of all ATA-32 baselines.  Thus, the signal will primarily
be detected in the auto-correlations from each antenna and will be absent from the
cross-correlations.  
The weakness of the signal, its spatially diffuse nature, 
and the difficult interference
environment at 327 MHz make this experiment
an excellent candidate for the cross-correlation
subtraction method.  

\subsection{Epoch of Reionization}

The Universe went through a phase transition at a redshift $z>6$
from neutral gas to ionized
gas corresponding to the appearance of the first stars, quasars or
other objects which produced ultraviolet photons
\citep{1972A&A....20..189S}. 
Prior to this epoch of reionization (EOR), all hydrogen was in an 
atomic or molecular state.  The spin flip transition of atomic hydrogen
(HI) is expected to produce emission that is essentially continuous
at frequencies less than $\nu_{\rm HI}/z_{\rm EOR}$, where $\nu_{\rm HI}$ is
the rest frequency of the HI transition and $z_{\rm EOR}$ is the
redshift of the EOR.  At frequencies above this critical frequency,
the signal disappears due to the ionization of all atomic hydrogen.
The expected amplitude of the signal is on the order of 10 mK, or $>30$ 
dB below foreground noise.
The redshift of the EOR places the signature at frequencies where
interference is a critical problem for radio telescopes.  

The emission is expected to be globally distributed with a characteristic
scale length of $\sim 10$ arcmin 
\citep{2004ApJ...608..622Z}.
Detection of the global signature is the first step
towards characterization of the EOR.  The cross-correlation subtraction
method can be applied to this problem provided that baselines of the
interferometer are sufficiently long to resolve out the global EOR
signal.  In practice, this implies baselines of $3$ km for an 
attempted detection at a frequency of 100 MHz ($z_{\rm EOR} \sim 13$).
Integration times must be less than 10 seconds to prevent decorrelation
of the interfering signal on the long baselines.

Since such an experiment may have baselines on many intermediate baselines,
it may be desirable to determine the astronomical signal from more than
just the auto-correlation functions.  A set of cross-correlations from
short-baselines may also be included in the summed astronomical signal
or have a separate interference signal removed from them.

\subsection{Focal Plane Array Feeds}

Focal plane array feeds are increasingly important tools for radio astronomy
\citep{1996PASA...13..243S}.  These feeds place multiple receivers
in the focal plane of a single antenna.  Typically, they are designed only
to produce auto-correlations for each receiver.   If designed to produce
the full set or a partial set of cross-correlations, however, 
these array feeds could 
make use of the cross-correlation subtraction method to mitigate interference.
The technique may also be of use in eliminating bandpass ripples due to
solar interference.

\section{Conclusions}

We have described a new method for cancellation of interference in the power domain for the specific case in which the astronomy signal is apparent only in the auto-correlation signal.  
Provided that the RFI is not resolved by interferometer,
cross-correlation signals are used to determine a high interference to noise ratio reference signal.  Analytical results suggest that the method could prove powerful at removing very weak interference for arrays with $N_a$ greater than a few.  Simulations demonstrate the basic performance of the algorithm.  There are a few issues for further research.  Most important among these is testing the effect of small source contributions to the cross-correlations.  This method highlights a critical aspect of RFI mitigation research:  the techniques that work best are those that are best-suited to the interferer and to the scientific goal.

%% ------------------------------------------------------------------------ %%
%
%  REFERENCE LIST AND TEXT CITATIONS
%
%% ------------------------------------------------------------------------ %%

%\bibliographystyle{apj}
%\bibliography{rfi}

%% ------------------------------------------------------------------------ %%
%
%  END ARTICLE (1/2)
%
%% ------------------------------------------------------------------------ %%
% PLEASE PLACE END ARTICLE AND NEW PAGE COMMANDS HERE FOR DRAFT MODE
% FOR GALLEY MODE REMEMBER TO (1) COMMENT OUT THESE LINES AND
% (2) PLACE AN END ARTICLE COMMAND AFTER THE FIGURES AND TABLES INSTEAD
\end{article}
\newpage

%% ------------------------------------------------------------------------ %%
%
% FIGURES (see end of examples for further instructions)
%
% ---------------
% Single column figure example
%

\end{document}